\documentclass{iaus}
\usepackage{graphicx}

\title[The Magnetosphere of $\sigma$ Ori E] 
{Modeling the Magnetosphere of the B2Vp star $\sigma$ Ori E}

\author[M. E. Oksala et al.]   
{M.E. Oksala$^{1,2}$, G.A. Wade$^2$, R.H.D. Townsend$^{3}$, \\ O. Kochukhov$^{4}$, S. Owocki$^{5}$}

\affiliation{$^{1}$Department of Physics and Astronomy,  University of Delaware, Newark, DE, USA\\[\affilskip]
$^{2}$Department of Physics, Royal Military College of Canada, Kingston, Ontario, Canada\\[\affilskip]
$^{3}$Department of Astronomy, University of Wisconsin-Madison, Madison, WI, USA\\[\affilskip]
$^{4}$Department of Physics and Astronomy, Uppsala University, Uppsala, Sweden\\[\affilskip]
$^{5}$Bartol Research Institute, University of Delaware, Newark, DE, USA}

\pubyear{2010}
\volume{272}  
\pagerange{1--2}
\jname{Active OB Stars: structure, evolution, mass-loss and critical limits}
\editors{C. Neiner, G. Wade, G. Meynet \& G. Peters, eds.}
\begin{document}

\maketitle

\begin{abstract}
This paper presents results obtained from Stokes I and V spectra of the B2Vp star sigma Ori E, observed by both the Narval and ESPaDOnS spectropolarimeters. Using Least-Squares Deconvolution, we investigate the longitudinal magnetic field at the current epoch, including period analysis exploiting current and historical data. $\sigma$ Ori E is the prototypical helium-strong star that has been shown to harbor a strong magnetic field, as well as a magnetosphere, consisting of two clouds of plasma forced by magnetic and centrifugal forces to co-rotate with the star on its 1.19 day period. The Rigidly Rotating Magnetosphere (RRM) model of Townsend \& Owocki (2005) approximately reproduces the observed variations in longitudinal field strength, photometric brightness, H$\alpha$ emission, and various other observables. There are, however, small discrepancies between the observations and model in the photometric light curve, which we propose arise from inhomogeneous chemical abundances on the star's surface. Using Magnetic Doppler Imaging (MDI), future work will attempt to identify the contributions to the photometric variation due to abundance spots and due to circumstellar material.
\end{abstract}

\firstsection 

\section{Introduction}

The term ``magnetosphere'' was coined to describe ``the region in the vicinity of the earth in which the Earth's magnetic field dominates all dynamical processes'' (\cite[Gold 1959]{Gold_1959}).  The Earth's magnetic field is distorted by the solar wind, compressed on the side closest to the sun and pulled out on the opposite side into the magnetotail.  As observations improved, it was shown that each planet with a magnetic field also possessed a similar magnetosphere to the Earth's, all created by the solar wind's effect on the magnetic field.  Eventually, ``magnetosphere'' was expanded to include stellar environments where winds and magnetic fields interact, e.g. the solar win and magnetic field.

Unlike planetary magnetospheres, stellar magnetospheres expand outward, as the wind is coming directly from the star itself.  We can understand solar system magnetospheres in great detail, describing the complex plasma physics, as we are able to obtain direct observations by satellites.  In the stellar case, we must use remote observations to infer physical properties (i.e., temperature, density), augmented by MagnetoHydroDynmaical (MHD) simulations.

There is a distinctive difference between solar and massive-star magnetospheres.  In lower mass stars like the sun, an $\alpha$-$\Omega$ dynamo combines a convective envelope with differential rotation to produce complex, small-scale, time variable field structures.  Globally, the field is quite weak.  The solar wind is pressure-driven with a relatively weak mass loss rate ($\sim$ 10$^{-14}$ M$_{\odot}$/yr).  More massive stars have mostly radiative envelopes.  This major difference makes magnetic field generation by a dynamo difficult to explain, so the question of the origin of magnetic fields in massive stars remains open.  Massive-star fields are also quite different from their solar counterparts.  Their fields are generally steady, large scale, and sometimes very strong.  Massive-star winds are radiatively driven with strong mass loss rates (on average $\sim$ 10$^{-7}$ M$_{\odot}$/yr).  Hereafter, this paper will focus solely on massive-star magnetospheres, specifically in the strong magnetic field case. 

\section{Massive star magnetospheres}

As the wind interacts with the magnetic field, material from the stellar wind can be perturbed, channeled, torqued, and even confined.  Massive star magnetospheres are generally structured and dynamic.  Material is released in reconnection events, falls back onto the star and creates collisional shocks.  Rotation complicates the physical picture, but it also allows for regular modulation of observables, through which the physical properties of these magnetospheres can be identified.  The magnetospheric properties depend on the relative strengths of the wind and magnetic energy densities of the star.  \cite[ud-Doula \& Owocki (2002)]{ud-Doula_Owocki2002} define a wind confinement parameter to quantify this struggle between the wind strength and magnetic strength, 
\begin{equation}
\eta_{\star} = \frac{B_{\star}^{2} R_{\star}^{2}}{\dot{M} v_{\infty}}.
\end{equation}
For $\eta_{\star} < 1$, the wind overwhelms the field, but for $\eta_{\star} > 1$, the magnetic field is dominant near the star, channeling wind material into a magnetosphere.  The Alfv\'en radius, R$_{\rm{A}}$, is where the magnetic and wind energy densities become equal.  For a strong magnetic confinement ($\eta_{\star} \gg 1$) and a dipole field configuration, R$_{\rm{A}} \sim \eta_{\star}^{-1/4} R_{\star}$ (\cite[ud-Doula \& Owocki 2002]{ud-Doula_Owocki2002}).  \cite[ud-Doula et al. (2008)]{ud-Doula_etal2008} also define a Kepler co-rotation radius (R$_{\rm{K}}$ = (v$_{rot}$/v$_{crit}$)$^{-2/3}$ R$_{\star}$).  If the Alfv\'en radius is father out than the Kepler radius a region of confinement is created.  Wind material is spun up by the magnetic field and supported by centrifugal forces.  Above R$_{\rm{A}}$, the field is too weak to confine material; below R$_{\rm{K}}$, it is no longer supported by centrifugal forces and so falls back on the star.

\section{The RRM model and $\sigma$ Ori E}

In many magnetic Bp stars, the magnetic field is strong, rotation is fast, and $\eta_{\star} \gg 1$.  The field spins up and confines the channeled wind keeping it  rigidly rotating well beyond the Kepler co-rotation radius, but also held down against  the net outward centrifugal force relative to gravity.  For this strong field limit, \cite[Townsend \& Owocki (2005)]{Townsend_Owocki2005} developed an analytical model, the Rigidly Rotating Magnetosphere (RRM) model, wherein wind plasma settles at the local minima of the effective (gravitational + centrifugal) potential.  The plasma then accumulates in the magnetosphere and is forced to co-rotate with the star.  

\begin{figure}[ht]
\begin{center}
 \includegraphics[width=4.3in]{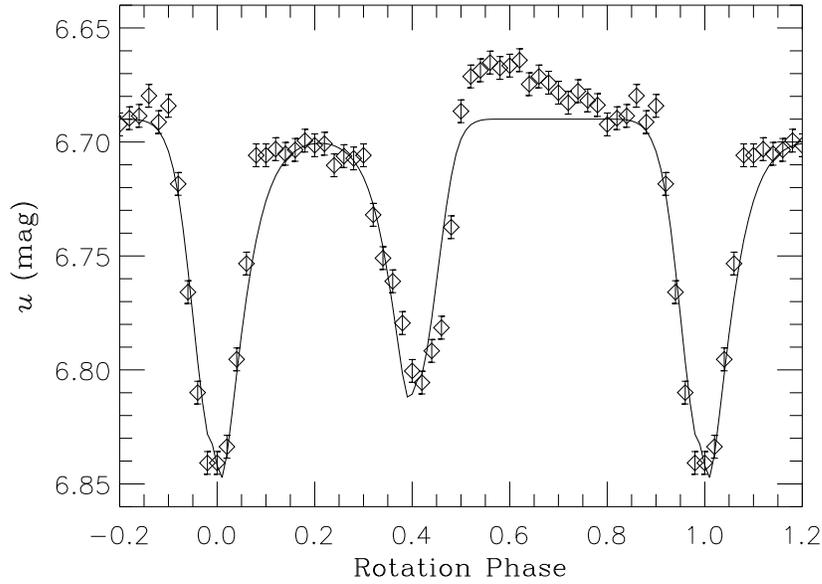} 
 \caption{Figure from \cite[Townsend et al. (2005)]{Townsend_etal2005}. Observed (diamonds) and modeled (solid line) Str$\ddot{\rm{o}}$mgren u-band
light curves of $\sigma$ Ori E, phased on the star's 1.19 day rotation period.  }
   \label{fig1}
\end{center}
\end{figure}

$\sigma$ Ori E (HD 37479) is a helium-strong B2Vp star with a 10 kG global dipole magnetic field.  It is an oblique rotator with two plasma clouds rigidly rotating with the star (\cite[Landstreet \& Borra 1978]{Landstreet_Borra1978}) with rotational speeds greater than the surface speed of $v \sin i = 150$ km s$^{-1}$.  Observations show modulation according to the 1.19 d rotation period in longitudinal magnetic field (\cite[Landstreet \& Borra 1978]{Landstreet_Borra1978}), H$\alpha$ emission (\cite[Walborn 1974]{Walborn_1974}), He line strength (\cite[Pedersen \& Thomsen 1974]{Pedersen_Thomsen1974}), photometry (\cite[Hesser et al. 1976]{Hesser_etal1976}), UV line strength (\cite[Smith \& Groote 2001]{Smith_Groote2001}), 6 cm radio emission (\cite[Leone \& Umana 1993]{Leone_Umana1993}), and linear polarization (\cite[Kemp \& Herman 1977]{Kemp_Herman1977}).  $\sigma$ Ori E also has anomalous surface abundances of He and Si (\cite[Reiners et al. 2000]{Reiners_etal2000}).  The RRM model reproduces the photometric, magnetic field, and H$\alpha$ observations relatively well.  It only accounts for the circumstellar effects that create the observed double eclipse pattern, but the stellar photosphere also contributes to the brightness variations.  This leads to discrepancies (Figure \ref{fig1}) between the observed and simulated photometric light curve, which we are currently investigating.

\section{Spectropolarimetric observations}

We obtained a total of 18 high resolution (R=65000) broadband (370-1040~nm) circular polarization spectra of $\sigma$ Ori E.  Sixteen of these spectra were obtained in November 2007 from the Narval spectropolarimeter attached to the 2.2m Bernard Lyot telescope at the Pic du Midi Observatory in France.  The remaining 2 spectra were obtained in February 2009 from the spectropolarimeter ESPaDOnS attached to the 3.6-m Canada-France-Hawaii Telescope, as part of the Magnetism in Massive Stars (MiMeS) Large Program (Wade et al., these proceedings).   

We used the method of Least-Squares Deconvolution (LSD) to derive values of longitudinal field for each spectrum.  LSD describes the stellar spectrum as the convolution of a mean Stokes I or V profile, representative of the average shape of the line profile, and a line mask, describing the position, strength and magnetic sensitivity of all lines in the spectrum.  A mask was used for a star with  $T_{\rm{eff}}$=23000~K, log $g$ = 4.0, and solar abundances except for enhanced helium.  From the LSD mean Stokes I and V profiles, we calculate the longitudinal magnetic field, B$_{\ell}$:
\begin{equation}
B_{\ell} = -2.14 \times 10^{11} \frac{ \int v V(v) dv}{\lambda g c \int [1-I(v)]dv}
\end{equation}
 (\cite[Wade et al. 2000]{Wade_etal2000}), where $\lambda$ is the weighted average wavelength and g is the weighted average Land\'e factor in the mask. I$_{c}$ is
the continuum value of the intensity profile. The integral is evaluated over the full velocity range of the mean profile. 

\begin{figure}[ht]
\begin{center}
 \includegraphics[width=5.0in]{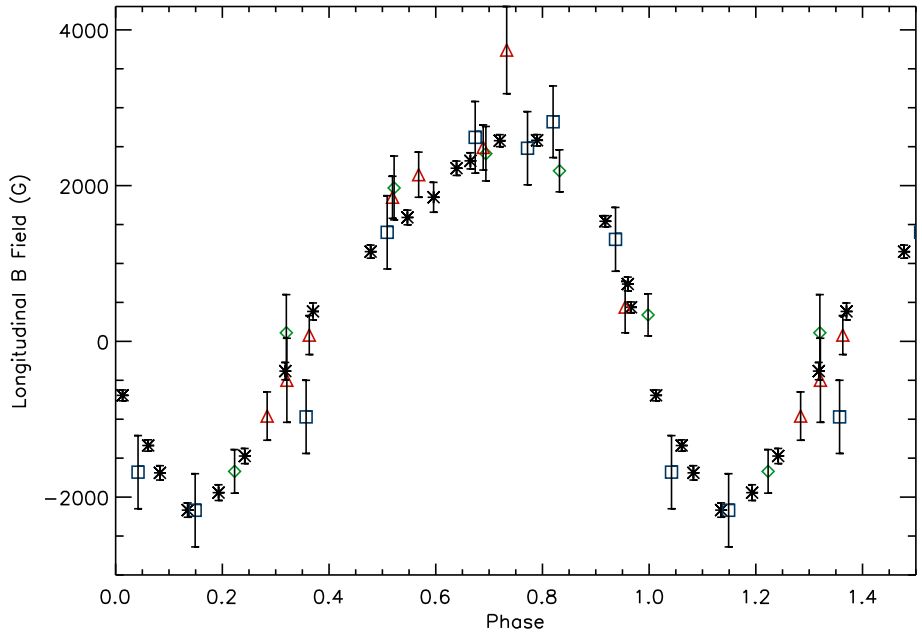} 
 \caption{New longitudinal field measurements for each individual spectrum of $\sigma$
Ori E (asterisks) along with 1$\sigma$ error bars. In addition, we have plotted measurements
from \cite[Bohlender et al. (1987)]{Bohlender_etal1987}, as triangles (hydrogen) and diamonds (helium), and \cite[Landstreet \& Borra (1978)]{Landstreet_Borra1978}
(squares) each with 1$\sigma$ error bars. }
   \label{fig2}
\end{center}
\end{figure}

When compared with historical data from \cite[Landstreet \& Borra (1978)]{Landstreet_Borra1978} and \cite[Bohlender et al. (1987)]{Bohlender_etal1987}, our new measurements agree surprisingly well, indicating stability of the global magnetic structure over three decades.  The new measurements are also much more precise, decreasing the error bars by a factor of four versus the historical data.  The spectropolarimetric data presented in Figure \ref{fig2} allow a long baseline to study the periodicity of the field as compared with the period derived from photometric studies.  Using the Scargle periodogram, the entire set of magnetic data gives a period of 1.190842 $\pm$ 0.000004 days.  The baseline of the data is long enough that the period can be determined down to a third of a second.  This is strictly an average period over the time frame of the observations.  As \cite[Townsend et al. (2010)]{Townsend_etal2010} have shown, $\sigma$ Ori E is spinning down due to magnetic braking at a rate of 77 milli-seconds per year.  The associated spindown time calculated from observations, 1.34 Myr, agrees remarkably well with the theoretical prediction of 1.4 Myr by \cite[ud-Doula et al. (2009)]{ud-Doula_etal2009}.

\section{Magnetic Doppler Imaging}

From rotationally modulated line profiles, the surface abundance and magnetic field vectors of a star can be mapped using a technique called Magnetic Doppler Imaging (MDI), originally developed by \cite[Piskunov \& Kochukhov (2002)]{Piskunov_Kochukhov2002} for the Ap star $\alpha^{2}$ CVn.  MDI was adapted from Doppler Imaging (DI), which reconstructs features on the surfaces of stars by inverting a time series of high-resolution spectral line profiles.  The code compares observational Stokes parameters with synthetic parameters, using least squares minimization.  MDI solves the radiative transfer equation to obtain the synthetic line profiles, assuming an unpolarized continuum and LTE.  The code can be used for arbitrarily complex fields and abundance distributions.

\begin{figure}[ht]
\begin{center}
 \includegraphics[width=5.0in]{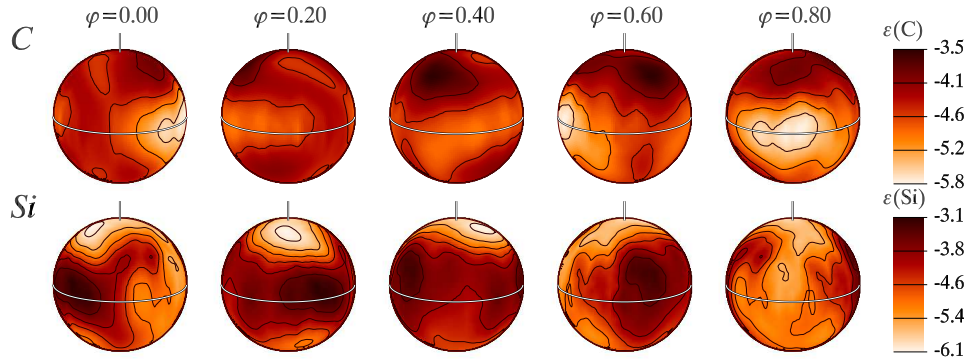} 
 \caption{The chemical abundance distribution of $\sigma$ Ori E derived from Stokes I and V profiles of the C and SI lines.  The star is shown at 5 equidistant rotational phases viewed at the inclination angle, $i=75^{\circ}$ and $v \sin i =150$ km s$^{-1}$. The scale gives abundance as $\varepsilon$(Elem) corresponding to $\log(N_{Elem}/N_{tot}$).  The rotation axis is vertical.}
   \label{fig3}
\end{center}
\end{figure}

This paper applies MDI to the Bp star $\sigma$ Ori E.  The maps created from circularly polarized line profiles  show a dipole magnetic field with a polar strength of 9.6 kG, although higher-pole magnetic field geometries and small scale fields cannot be identified without Stokes Q and U information.  The surface of the star (as shown in Figure \ref{fig3}) shows spots in carbon and silicon.   The helium lines of $\sigma$ Ori E show NLTE effects, especially at longer wavelengths.  Currently, MDI is fitting helium lines with a partial NLTE implementation to include departure coefficients calculated from the NLTE atmosphere code TLUSTY (\cite[Lanz \& Hubeny 2007]{Lanz_Hubeny2007}).  It may be possible to fit other elements from $\sigma$ Ori E's spectrum, however rapid rotation and lack of available lines affect MDI's ability to calculate accurate maps.

\section{Synthetic photometry}
 Krti\u{c}ka et al. (2007) used DI maps of HD 37776 to study the effects of spectroscopic spots on the surface of a star on the redistribution of flux.  The maps of silicon and helium used show that both elements contribute to changes in the continuum of the star's emergent flux.  In their paper, LTE atmospheres are used to calculate radiative fluxes and simulate light curves for each element.  The authors note that NLTE effects in lines may change the shape of the predicted light curve.  Matching the observed photometric light curve requires a combination of each elemental light curve.  This method has been shown to be quite successful at reproducing the light curve of HD~37776, as the star does not show an eclipse pattern expected from a magnetosphere.   
 
For other stars like $\sigma$ Ori E with magnetospheres, Magnetic Doppler Imaging (MDI) can be used in the same way to create synthetic light curves for specific elements, taking into account the effects of chemical spots on the total flux emitted due to bound-free transitions.  In \S 3, we mentioned that the RRM model can provide a synthetic light curve for the circumstellar effects on brightness.  To fully account for the contributions to the emergent flux from $\sigma$ Ori E, we should merge both the light curves synthesized from the RRM model and from MDI.  Ideally, when we combine both the photospheric and circumstellar brightness variations, we should be able reproduce the details of the observed photometric light curve.

\section{Summary}

Massive star magnetospheres are structured, co-rotate with the star, and produce rotationally modulated observables.  The RRM model of the circumstellar environment of $\sigma$ Ori E can reproduce observations relatively well, except for unexplained differences in the photometric light curve.  From new spectropolarimetric observations we show that $\sigma$ Ori E has a strong (9.6 kG), dipole magnetic field that new observations show has remained steady in structure over 3 decades.  Its longitudinal magnetic field varies from -2.3 kG to 2.5 kG.  The period derived from these longitudinal field strength variations is 1.190842 $\pm$ 0.000004 days.  This precise period is the average over the time baseline, as $\sigma$ Ori E's rotation period gains 77 milli-seconds  each year due to magnetic braking.  We can use MDI to map the surface abundance of $\sigma$ Ori E and synthesize a photospheric brightness light curve.  By merging the light curves from both the MDI and RRM models, we expect to properly account for both the circumstellar and photospheric effects on the star's brightness.  Using this combined approach, we hope to reproduce the details of the observed photometric light curve of $\sigma$ Ori E. 

\section{Acknowledgments}

MEO and RHDT acknowledge support from NASA grant grant {\it LTSA}/NNG05GC36G.  O.K. is a Royal Swedish Academy of Sciences Research Fellow supported by grants
from the Knut and Alice Wallenberg Foundation and the Swedish Research Council.


\begin{thebibliography}{}

\bibitem[Bohlender \etal\ (1987)]{Bohlender_etal1987}
{Bohlender, D.A., Landstreet, J.D., Brown, D.N., Thompson, I.B.} 1987,
\textit{ApJ}, 323, 325 

\bibitem[Gold (1959)]{Gold_1959}
{Gold, T.} 1959,
\textit{JGR}, 64, 1665 

\bibitem[Krticka \etal\ (2007)]{Krticka_etal2007}
{Krti\u{c}ka, J., Mikul\'a\u{s}ek, Z., Zverko, J., \u{Z}i\u{z}\u{n}ovsk\'y} 2007,
\textit{A\&A}, 470, 1089

\bibitem[Hesser \etal\ (1976)]{Hesser_etal1976}
{Hesser, J.E., Walborn, N.R., Ugarte, P.P.} 1976,
\textit{Nature}, 262, 116 

\bibitem[Kemp \& Herman (1977)]{Kemp_Herman1977}
{Kemp, J.C. \& Herman, L.C.} 1977,
\textit{ApJ}, 218, 770 

\bibitem[Landstreet \& Borra (1978)]{Landstreet_Borra1978}
{Landstreet, J.D. \& Borra, E.F.} 1978,
\textit{ApJ} (Letters), 224, 5 

\bibitem[Lanz \& Hubeny (2007)]{Lanz_Hubeny2007}
{Lanz, T \& Hubeny, I.} 2007,
\textit{ApJS}, 169, 83

\bibitem[Leone \& Umana (1993)]{Leone_Umana1993}
{Leone, F. \& Umana, G.} 1993,
\textit{A\&A}, 268, 667 

\bibitem[Pedersen & Thomsen (1977)]{Pedersen_Thomsen1977}
{Pedersen, H. \& Thomsen, B.} 1977,
\textit{A\&AS}, 30, 11 

\bibitem[Piskunov \& Kochukhov (2002)]{Piskunov_Kochuvhov2002}
{Piskunov, N. \& Kochukhov, O.} 2002,
\textit{A\&A}, 736, 756

\bibitem[Reiners et al. 2000]{Reiners_etal2000}
{Reiners, A., Stahl, O., Wolf, B., Kaufer, A., Rivinius, T.} 2000,
\textit{A\&A}, 363, 585

\bibitem[Smith & Groote (2001)]{Smith_Groote2001}
{Smith, M.A. \& Groote, D.} 2001,
\textit{A\&A}, 372, 208 

\bibitem[Townsend \& Owocki (2005)]{Townsend_Owocki2002}
{Townsend, R.H.D. \& Owocki, S.P.} 2005,
\textit{ApJ}, 576, 413 

\bibitem[Townsend \etal\ (2005)]{Townsend_etal2005}
{Townsend, R.H.D., Owocki, S.P., Groote, D.} 2005,
\textit{ApJ} (Letters), 630, 81

\bibitem[Townsend \etal\ (2010)]{Townsend_etal2010}
{Townsend, R.H.D., Oksala, M.E., Cohen, D.H., et al.} 2010,
\textit{ApJ} (Letters), 714, 318

\bibitem[ud-Doula \& Owocki (2002)]{ud-Doula_Owocki2002}
{ud-Doula, A. \& Owocki, S.P.} 2002,
\textit{ApJ}, 576, 413 

\bibitem[ud-Doula \etal\ (2008)]{ud-Doula_etal2008}
{ud-Doula, A., Owocki, S.P., Townsend, R.H.D.} 2008,
\textit{MNRAS}, 385, 97 

\bibitem[ud-Doula et al. (2009)]{ud-Doula_etal2009}
{ud-Doula, A., Owocki, S.P., Townsend, R.H.D.} 2009,
\textit{MNRAS}, 392, 1022

\bibitem[Wade \etal\ (2000)]{Wade_etal2000}
{Wade G. A., Donati J.-F., Landstreet J. D., Shorlin S. L. S.} 2000, 
\textit{MNRAS}, 313, 851

\bibitem[Walborn (1974)]{Walborn1974}
{Walborn, N.R.} 1974,
\textit{ApJ} (Letters), 191, 95


\end{thebibliography}
\end{document}